\documentclass[iop,tighten]{emulateapj}

\newcommand{\msun}{M_\odot}

\usepackage{aas_macros}
\usepackage{units}
\def\msun{M_{\odot}}

\begin{document}

\title{The formation of Milky Way-mass disk galaxies in the first 500 
million years of a cold dark matter universe}
\shorttitle{Formation of first Milky Way-mass disk galaxies}
\shortauthors{Feng et al.}

\author{Yu Feng\altaffilmark{1},
\and
{Tiziana Di Matteo\altaffilmark{2}},
\and
{Rupert Croft\altaffilmark{2}},
\and
{Ananth Tenneti\altaffilmark{2}},
\and
{Simeon Bird\altaffilmark{2}},
\and
{Nicholas Battaglia\altaffilmark{3}},
\and
{Stephen Wilkins\altaffilmark{4}}
}
\email{Email: yfeng1@berkeley.edu}

\altaffiltext{1}{Berkeley Center for Cosmological Physics, University of California
at Berkeley, Berkeley, CA 94720, USA}
\altaffiltext{2}{McWilliams Center for Cosmology, Physics Dept., Carnegie Mellon 
University, Pittsburgh PA 15213, USA}
\altaffiltext{3}{Department of Astrophysical Sciences, Princeton University,
Princeton, NJ 08544, USA}
\altaffiltext{4}{Astronomy Center, Dept. of Physics and Astronomy, University
of Sussex, Brighton, BN19QH, UK}

\begin{abstract}
Whether among the myriad
tiny proto-galaxies there exists a population with similarities
to present day galaxies is an open question. 
We show, using BlueTides,
the first hydrodynamic simulation large enough
to resolve the relevant scales, that the first massive galaxies to form
are 
predicted to have extensive rotationally-supported disks. Although their morphology resembles in some ways Milky-way types seen at much lower redshifts, these high-redshift galaxies are smaller, denser, and richer in gas than their low redshift counterparts.
From a kinematic analysis of a statistical sample
of 216 galaxies at redshift $z=8-10$
we have found that disk galaxies make up 70\% of the population of
galaxies with stellar mass $10^{10} M_\odot$ or greater. 
Cold Dark Matter cosmology therefore makes specific predictions for
the population of large galaxies 500 million years after the Big Bang.
We argue that wide-field satellite telescopes (e.g. WFIRST) will in the near
future discover these first massive disk galaxies.
The simplicity of their structure and formation history should make possible new tests of cosmology.

\end{abstract}

\maketitle

\section{Introduction}
\begin{figure*}[ht]
\includegraphics[width=\textwidth]{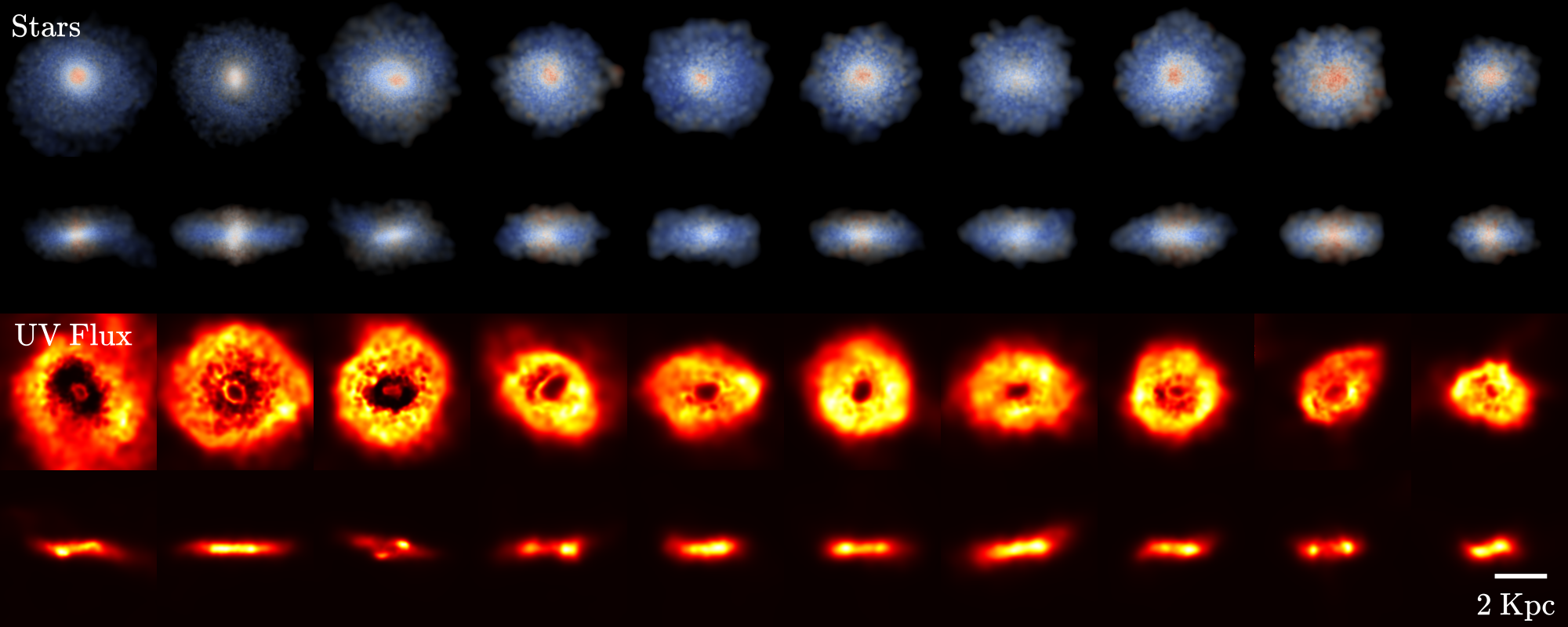}
\caption{
A sample of disk galaxies selected from the BlueTides simulation at redshift
$z=8$. We show galaxies identified kinematically
to be rotating disks. 
For each galaxy we show a face-on and 
side-on view. 
The colors of top two rows represent 
the stellar surface density 
coded by  the stellar age (blue to red: $0\sim~300\,\mathrm{Myear}$), face-on and side-on.
The colors of the bottom two rows represent the star-formation surface density (normalized to $0 \sim 1$). The face-on images have included the effect of dust-extinction. The darker central regions of the disks are mostly caused by the consumption of gas to form bulge stars. There is also an obscuration effect from the higher metallicity (and therefore dust content) of the central regions.}
\end{figure*}

The search for ``primeval'' galaxies and the understanding 
that will come from studying them has long been a guiding
principle for much of extra-galactic astrophysics and cosmological research.
Theories for the formation of the first galaxies span early models of monolithic
collapse of massive bodies\citep{eggen62} to merger scenarios for
the hierarchical formation of all structures\citep{press74}. Disk galaxies are arguably
those that have generated the most interest, and were the focus of early theoretical work\citep{fall80}. 
In the simplest galaxies, the structure of rotating disks should be directly 
linked to the angular momentum of the surrounding dark matter halo. Finding 
galaxies for which this is true, in the earliest
environments possible would enable powerful tests of the dynamics,
gaseous and radiative processes involved in galaxy formation.

In current cosmological models, galaxies form from the 
gravitational collapse of small perturbations in the matter 
distribution\citep{silk12}.
This process involves both a hierarchy of 
merging structures and smooth accretion, so that early galaxies are
predicted to be morphologically irregular, clumpy, and 
compact\citep{shimizu14,cen14}.
This is supported by recent
observational data on samples of galaxies at redshift $z=8$
and beyond\citep{oesch10,ono13,curtis14}. The volumes accessible to these studies, both
computational and observational are however thousands of times smaller
than those that will be probed by upcoming telescopes\citep{wfirst}.

The earliest that large rotating disks have been seen observationally
is at redshifts $z=2-3$ \citep{genzel06,glazebrook13}, and this has required deep spectroscopy
on 10m-class telescopes. At higher redshifts, galaxies in published
samples are very compact, with half light radii $ \sim 0.5$ kpc at $z=8$\citep{oesch10,holwerda14}. The fraction of morphologically disturbed galaxies
is also known to be high in deep fields
\citep{menanteau06}. The volumes probed at redshifts $z>3$
and particularly the redshifts at the current frontier, $z=8-10$
are however very small, with a sky areas of 11 sq. arcmin accessible
to the Hubble Extreme Deep Field\citep{illingworth13}
for example. This has meant that probing the 
highest luminosity and largest galaxies, has not been possible so far.
It is not surprising that observational studies of the structure and morphologies
of the highest $z$ galaxies have not revealed large disk galaxies, whether they
exist or not.

Theoretically, the situation is similar - it is not known what the
prevailing Cold Dark Matter model predicts for the most massive galaxies
at redshifts $z\geq8$. 
Numerical simulations used to make predictions have so far\citep{jaacks12,cen14} been limited
to  volumes 40-100 times smaller than our present work, 
or else zoomed simulations concentrating on individual 
galaxies\citep{pawlik11} or overdensities\citep{yajimal15}. Intriguingly, a dominant fraction of
the early forming (dwarf-mass) galaxies
simulated so far show evidence of disk-like structure\citep{romanodiaz11,pawlik11}.
It is of prime importance to reach the regime of large galaxies as the 
early formation of luminous or massive objects (such as first bright quasars, and most massive galaxy clusters) is one
of the most potent tests of models of structure formation in hierarchical cosmology.
In order to do so, and simulate a statistically useful sample
of galaxies, it
is necessary to model a volume of order several hundred megaparsecs on a side.
This must be done with high enough mass and force resolution to allow
studies of structure and morphology (See e.g., \cite{governato07}).
We have been able to reach this regime, with the BlueTides run,
a cosmological hydrodynamic simulation which we have just completed, and
we describe the first results below.

\section{The BlueTides simulation}
BlueTides was carried out
using the Smoothed Particle Hydrodynamics code MP-Gadget (see \cite{springel05a,dimatteo12}
for earlier versions of the code)
with $2\times7040^{3}$ (0.69 trillion) particles on 
648,000 Cray XE compute cores of the Blue Waters system at the 
National Center for Supercomputing Applications (NCSA). The simulation
evolved a cube of side length 400 comoving Mpc/h to redshift $z=8$,
and is the largest cosmological hydrodynamic simulation yet carried out
in terms of memory usage,
by an order of magnitude\citep{dimatteo12}.
The supernova feedback in BlueTides is modeled after \cite{okamoto10}, with the wind feedback parameter $\kappa_w=3.7$. This model has been shown to reproduce several key properties of low redshift galaxies \citep{illustris}. We also include the effect of super-massive black holes,  primordial and metal cooling, H2 molecular cloud and a non-uniform ionization background: we refer to \cite{reionpaper} for the parameters and
other aspects of the simulation in more detail. We also show there that the global star formation rate predicted is consistent with current observations at redshifts $z>8$.
The volume simulated is approximately 300 times larger than the largest 
observational survey at these redshifts\citep{trenti12}.

Galaxies were selected from the raw particle data (47 TB per output time)
using a friends-of-friends algorithm at a range of redshifts from $z=8-13$. The 
mass resolution of the simulation is $1.72\times 10^{7} M_\odot$ per dark matter particle, 
so that at redshift $z=8$,
36 million galaxies were found containing 64 dark matter particles or more, above
a mass threshold of $10^{9} M_\odot$. The BlueTides galaxy luminosity
functions at the redshifts of interest here are fully consistent with  current observations, and this is shown in \cite{reionpaper}.
In the present case, we are most
interested in the most massive galaxies with at least $10^4$ stellar particles, which contain enough particles to study their rotational support. The corresponding stellar mass is $0.7\times10^{10} \msun$ or more.

We have made images of all galaxies present in the
simulation,  by adaptively smoothing
the star particle distribution
to reveal as much structure as possible given the resolution available.
We have also used the star formation rate (SFR) of the gas particles in the 
simulation to make other images. The SFR can be related to the
rest-frame UV ($150\,\mathrm{nm}$) luminosity. The UV luminosity
is sensitive to the presence of dust, and there is evidence
that the highest luminosity early galaxies are significantly 
obscured\citep{wilkins13,cen14}. We account for this dust attenuation using a screening model\citep{joung09}. This simple model assumes that the 
dust attenuation $A_{\rm UV}$ in a galaxy pixel is proportional to the
metal-mass density in that pixel. There is only one free parameter,
which we normalize by setting $A_{\rm UV}=1$ for observed $M_{\rm UV}=-21$
galaxies\citep{wilkins13}. We have taken this approach, rather than performing a more detailed radiative transfer simulation because the current major uncertainties lie in the dust content and attenuation curve\citep{wilkins13}.

\section{Galaxy Morphology}

We turn directly to the morphology of the most massive ($M_{\rm stellar} \ge 0.7\times10^{10} \msun$) galaxies. 
We compute the inertia tensor of the
star particles and use these to plot views of each galaxy parallel and
perpendicular to the major axis.
Studying them reveals that a significant fraction are
 visually disk-like, and surprisingly regular in shape. 

In Figure 1 we show images of 10 disk galaxies in the simulation at $z=8$,
both side-on and face-on for each.
These are among those kinematically selected to be disks (see below).
We can see that the disk components extend to a diameter of $\sim$ 4 kpc, with a median
half-light radii $\sim$0.6 kpc for those shown.
Although the $z=8$ BlueTides disk galaxies have masses comparable to
the Milky Way, their sizes are significantly smaller, by a factor 
of $\sim 5$. The disk galaxies also have mean gas fractions of  $\sim 70\%$
at $z=8$. 
The typical stellar mass/halo mass ratio of these systems is a few percent ($\sim 5\%$, estimated from Friend-of-Friend halos). This is similar to systems at z=0, where the  stellar to halo ratio peaks at 4\% for halos in the same mass range \citep{guo10}.  The star formation rates of these galaxies are  $\sim 100 \,\unit{M_\odot / year}$, orders of magnitude larger than the main sequence galaxies of similar stellar mass at low-z \citep{whitaker12}.
The stellar populations in the central regions of each galaxy
are significantly older than the outer disk, with a proportion of stars
that formed as early as $z=13$, i.e. age $300$ Myr at $z=8$.

These disk galaxies host  supermassive blackholes, with blackhole masses ranging from a few times $10^6$ to a few times $10^8$ $h^{-1}\mathrm{M}_\odot$.  
Interestingly, the host galaxy of the most massive blackhole ($3\times 10^8 h^{-1} \mathrm{M}_\odot$) has a much more dominate spheroidal component ($D/T = 0.29$) than typical galaxies with similar  masses ($D/T > 0.5$). We plan to investigate the role of central blackholes in the formation of disks in the near future.

\begin{figure}
\includegraphics[width=\columnwidth]{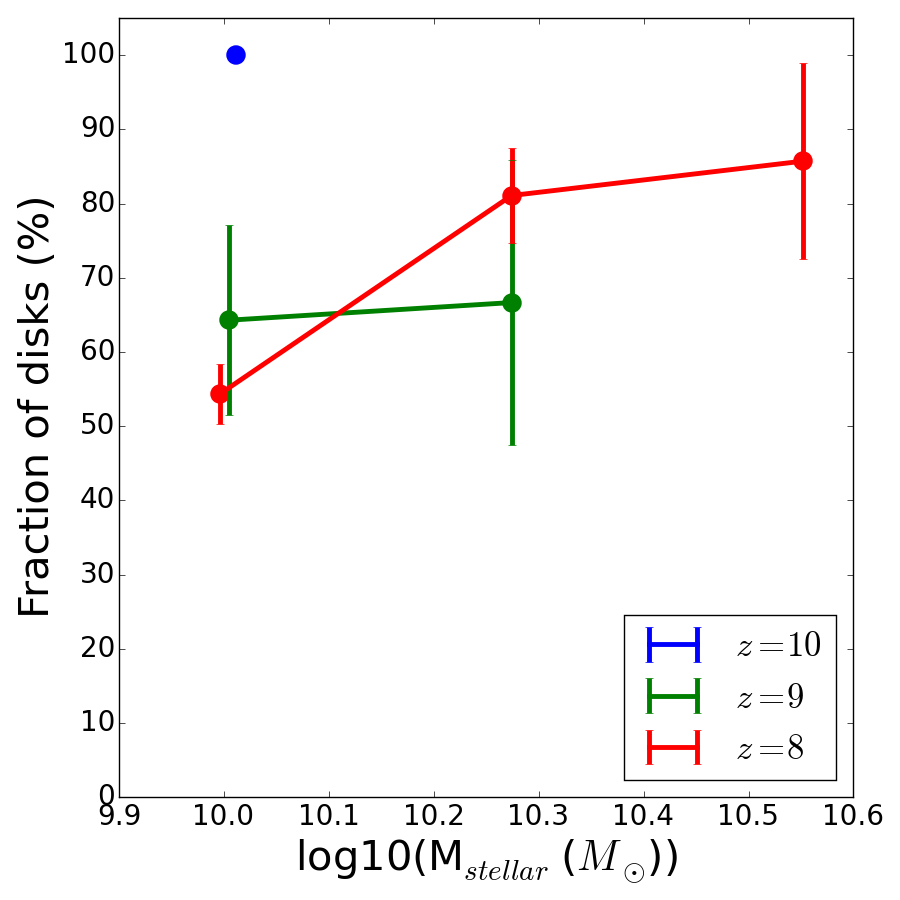}
\caption{
Fraction of galaxies in the BlueTides simulation kinematically classified as disks. Results are shown as a function of
galaxy stellar mass, and at 3 different redshifts.}
\end{figure}

\begin{figure}
\includegraphics[width=\columnwidth]{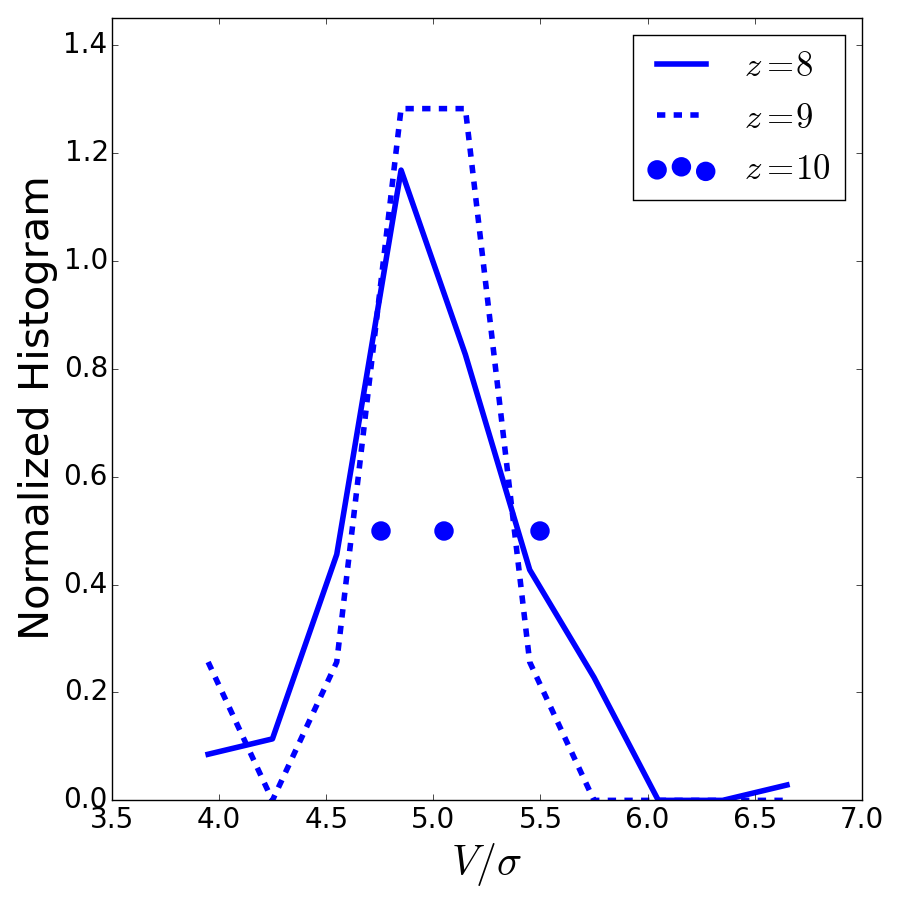}
\caption{
Disk galaxy kinematics in BlueTides characterized by the ratio of circular velocity $V$ to vertical velocity dispersion $\sigma$. We show histograms of results for this $V/\sigma$ ratio computed using star particles in the disks of galaxies selected to be disks from their D/T ratio. Data on redshifts $z=8$ and $z=9$ are shown.}
\end{figure}

Kinematic decomposition of the galaxies lets us know if the disk-like
structures seen in Figure 1 are rapidly rotating. We have used the standard
technique\citep{governato07} to determine the fraction of stars in each galaxy
which are on planar circular
orbits, and which are associated with a bulge. We note that this technique is not sensitive to the clumpiness of disks. At redshift $z=8$ 
we find that 70\% of
galaxies above a mass of $10^{10}\msun$ are kinematically classified 
as disks (using the standard threshold of disk stars to total stars
(D/T) ratio of 0.2\citep{governato07}. This percentage of disks changes with
redshift as shown in Figure 2. In all cases it is much higher than the
fraction of observed galaxies of comparable mass which are seen to
be disk-like at $z=0$, $0.14\pm0.03 $\citep{buitrago13}.
 
We note that the majority of gas-rich disks seen at high redshifts accessible so far to spectroscopy (z=2-3) are clumpy, turbulent and thick \citep{forster09}
(\cite{genzel06} is an exception), and it is
natural to expect this to be the case with the BlueTides massive disks, which have high gas fractions. A relevant statistic we have computed is the ratio of circular velocity, $V$ to vertical velocity dispersion $\sigma$ of our disk galaxies. We have computed this $V/\sigma$ ratio using information from the disk star particles and plot a histogram of the results in Figure 3. We can see that the values range from $V/\sigma=4$ to $6$, with a mean
value of $4.98 \pm 0.04$ measured from all redshifts. The galaxies are therefore definitely disks, having $V/\sigma$ well above the $0-1$ expected for non-disk galaxies. This supports our conclusions made using the D/T ratio above. The $V/\sigma$ values are lower than the $V/\sigma \sim 10$  expected for thin spiral galaxies seen at $z=0$, and instead consistent with gas rich turbulent disks from earlier times\citep{forster09}. Although this is reasonable, confirmation of this aspect awaits future higher resolution resimulations which are able to resolve the fine vertical structure of the disks.  

\section{Half-light radius}

\begin{figure}
\includegraphics[width=\columnwidth]{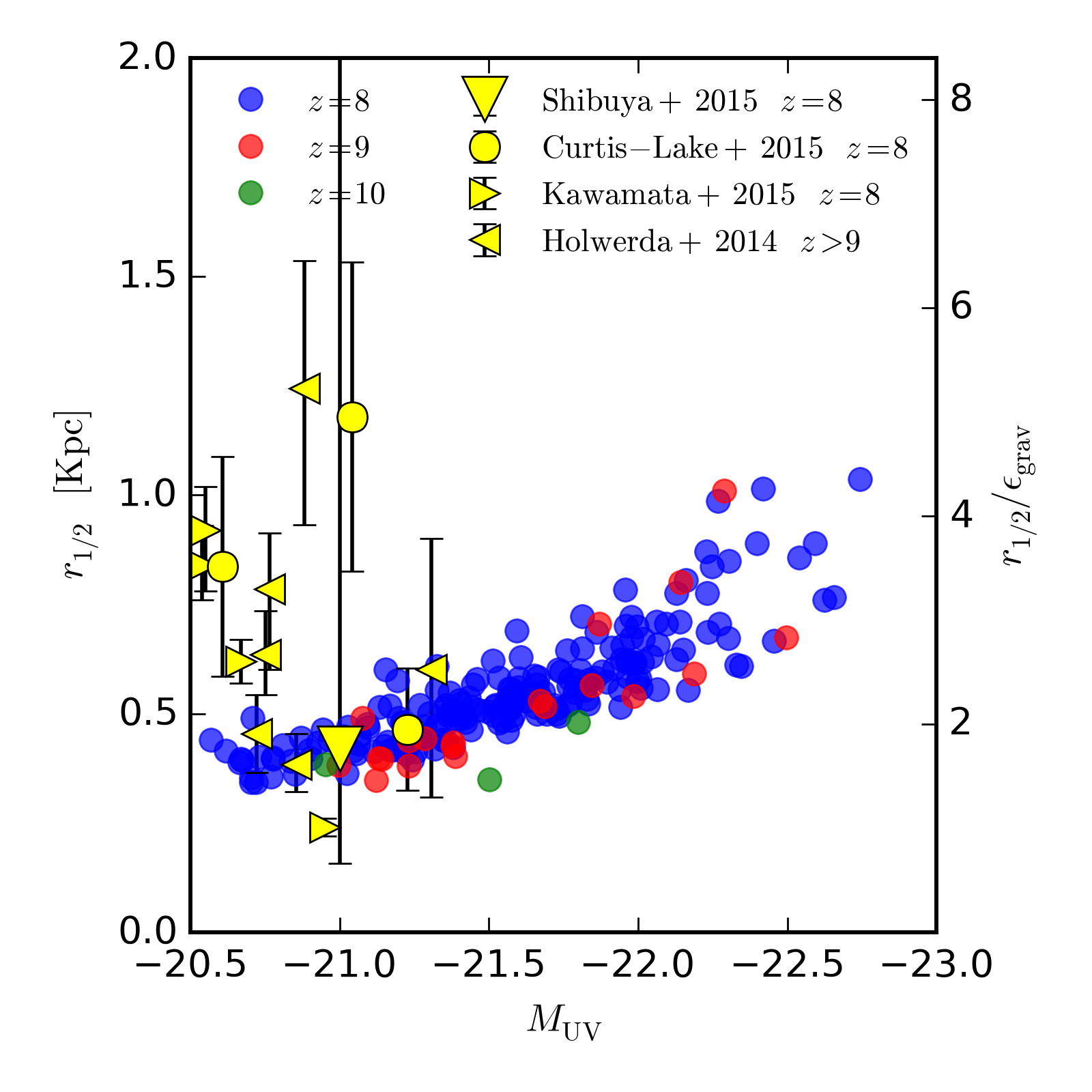}
\caption{
Galaxy half-light radii at redshift $z=8,9,10$ in the BlueTides
simulation (including disks and non-disks) and in the Hubble Space Telescope observations
of Holwerda et al (2014), Kawamata et al (2014), Curtis-Lake (2014), and most recently Shibuya et al (2015).
Colored Symbols: Galaxies in BlueTides Simulation; We include galaxies with a stellar mass $>5\times 10^{9}\,h^{-1}M_\odot$
Upward wedges: Holwerda et al (2014) $z \ge 9$ samples;
Downward wedges: Kawamata et al (2015) $z = 8$ samples.
Circles: Curtis-Lake et al (2014) $z = 8$ undisturbed samples $M_\mathrm{UV}>-20.5$, with uncertainty estimated from their Figure 1.
The half-light radius in BlueTides is measured from the face-on images. The right-axis is in units of the force resolution $\epsilon_\mathrm{gravi} = 0.24\,\mathrm{Kpc}$ at $z=8$.
}
\end{figure}

A number of recent studies\citep{oesch10,ono13,holwerda14,kawamata15,shibuya15} have begun to make measurements of galaxy
structure and in particular sizes at the highest redshifts. As the
observational volumes are small, the magnitude range of observed samples
varies from $M_{\rm UV}=-21$ to $M_{\rm UV}=-18$. We can therefore
compare BlueTides galaxies with these measurements, bearing in mind that the brightest galaxies in the simulation are much rarer than can be observed in current surveys.
Using the standard observational algorithm SEXtractor\footnote{We use a Python rewrite of SEXtractor by \cite{SEP}.}, we measure
the half-light radius ($r_{1/2}$)
 of each galaxy. This is shown in Figure 4, where we compare our
redshift $z=8$ results with observational
data from \citep{kawamata15}, \citep{holwerda14},
\citep{curtis14} and \citep{shibuya15}. We can see that the current observations cover the lower end of the magnitude range, and are in the same realm as (but not a good match to) our simulation results, with a mean $r_{1/2}\sim 0.6$
for $M_\mathrm{UV} \sim -20.5 \to -21.5$. 
Galaxies are therefore extremely compact at these high redshifts. 
The galaxy sizes in the simulation do appear to be on the low end
compared to observations, something which has also been seen at z=3\citep{joung09} and could be a sign of interesting physics. It has
also been shown however\citep{curtis14} that selection effects, flux cuts and choice of size measurement algorithms can affect measured galaxy
sizes by factors of 2 or more. 

The half light radius of the faintest samples ($\sim 0.5 \unit{Kpc}$) is only a factor of 2 of the gravitational smoothing length (0.24 Kpc at $z=8$), which could also affect our results. 
In future work we will explore these effects in more detail with mock observations.
Interestingly, the simulation predicts a positive size-luminosity correlation. 

Analytic models can be used to predict how disk sizes scale with redshift. For example, the simplest assumption is that the half-light radii scales with virial radius, which has the following redshift dependence 
\[
r_{vir}\propto H(z)^{-1} \propto (1+z)^{-1.5},
\]
for constant halo circular velocity (e.g., \cite{mo98}). 
Using this relationship to extrapolate the disk sizes from the GEMS observational sample analyzed in \cite{somerville08} (in which the median $r_d$ for galaxies with $m_{*}=10^{10} \unit{M_\odot}$ at $z=1.05$  is $2.1 \unit{Kpc}$), this simplest model predicts that disks of this mass should have $r_d = 0.23\ \unit{kpc}$ at $z=8$ and $r_d = 0.17\ \unit{kpc}$ at $z=10$. 
This is somewhat lower than disks of similar stellar mass in our simulation which have half-light radii about twice this value (see Figure 4 - these correspond to the least luminous disks plotted). \cite{somerville08} have made more sophisticated models which include evolution in the internal structure of dark halos and adiabatic contraction, both of which mean that the trend of size evolution with redshift is expected to be weaker than in the simplest models (also supported by observations, e.g. \cite{shibuya15}), and which therefore appear to describe our simulated results better.

\section{Conclusions}

At redshifts $z=8-10$ the Universe is about 1000 times denser than
the present day. In these very different physical conditions, and
in the very short time available, massive
disk galaxies are able to form in the cold dark matter model. They 
share many visual and kinematic characteristics with those that 
are seen when the Universe was 20 times older. This means that an image one might have of the early Universe containing only small protogalaxies, merging clumps and irregular structures is not appropriate.

An immediate question we can ask is how these disks did in fact form
in the simulation.
We have tracked particles in the most massive galaxies (including those plotted
in Figure 1) back in time to visually assess their formation history.
We have found that of the most massive 20 galaxies at redshift $z=8$,
only one was the product of an obvious major merger between similarly sized
progenitor galaxies. A popular  picture of gas rich major mergers producing some disks 
\citep{springel05} therefore does not appear to be relevant here.

Even though the Lagrangian volumes \citep{onorbe14} and halos \citep{rossi11} at the rare peaks associated with
these galaxies are likely more spherical,
 the distribution of matter around galaxies is known to be highly anisotropic in the CDM model, the infall of gas cannot proceed quasi-spherically as in the classical picture of galaxy formation\citep{fall80}. 
Instead, smooth infall 
and accretion build up the disks, with cold gas arriving along the directions
of nearby filaments in the density distribution \citep{romanodiaz14}.

These massive disks will be fascinating objects to study with 
next generation telescopes. The WFIRST satellite\citep{wfirst} will have a field of view 200 times that of the WFC3 instrument on HST, making a planned deep (J=26.7) sky survey covering 2000 square degrees possible. Using the luminosity function in BlueTides and its evolution we predict that this survey should find $\sim 8000$ disk galaxies of the type we have kinematically identified as massive disks, about 1 per field of view on average. This compares to a probability of 0.3 for finding a single one of these objects in the current largest area HST survey (BoRG, \cite{borg}).
 The rotation curves will be
exciting targets for Thirty Meter class ground based telescopes.
Forming from lightly enriched material, these primeval disk galaxies 
are special objects. Observing them and comparing them with 
simulations and analytic theory should enable us to make inferences
about dark matter, the role of angular momentum and the fundamental
principles of galaxy formation (such as tidal torques, and smooth accretion of unenriched material) which 
are not possible in the later, more complex Universe.

\acknowledgments
We acknowledge funding from NSF OCI-0749212
and NSF AST-1009781. The BlueTides simulation was run on BlueWaters supercomputer at  the National Center for Supercomputing Applications.

\bibliography{naturebib.bib}

\end{document}